\documentclass[preprint2]{aastex631}
\begin{document}

\title{A JWST study of CO$_2$ on the satellites of Saturn}

\correspondingauthor{Michael Brown}
\email{mbrown@caltech.edu}

\author[0000-0002-8255-0545]{Michael E. Brown}
 \affiliation{Division of Geological and Planetary Sciences, California Institute of Technology, Pasadena, CA 91125, USA}

\author[0000-0002-0767-8901]{Samantha K. Trumbo}
\affiliation{Department of Astronomy \& Astrophysics, University of California, San Diego, La Jolla, CA 92093, USA}

 \author[0000-0003-4778-6170]{Matthew Belyakov}
 \affiliation{Division of Geological and Planetary Sciences, California Institute of Technology, Pasadena, CA 91125, USA}

\author[0000-0002-7451-4704]{M. Ryleigh Davis}
\affiliation{Division of Geological and Planetary Sciences, California Institute of Technology, Pasadena, CA 91125, USA}
 
\author[0000-0003-3303-1009]{Ashma Pandya}
\affiliation{Division of Geological and Planetary Sciences, California Institute of Technology, Pasadena, CA 91125, USA}

\begin{abstract}

Solid state CO$_2$ has been detected throughout
the outer solar system, even at temperatures
where
crystalline CO$_2$ is unstable, requiring that
the CO$_2$ be trapped in a separate
host material. 
The Saturnian 
satellites 
provide an ideal laboratory for the study of 
this trapped CO$_2$,
allowing us 
to examine objects with 
identical insolation, but with a range of 
environments, ice exposure, organic abundance, 
and formation locations. Here, we present JWST spectra of
8 mid-sized satellites of Saturn, including Mimas, Enceladus, Tethys, Dione, and Rhea
interior to Titan, and Hyperion, Iapetus, and
Phoebe exterior. The $\sim$4.26~$\mu$m CO$_2$ $\nu_3$ band is detected on each 
satellite, and the $\sim$2.7~$\mu$m $\nu_1+\nu_3$ band
is detected on all but Phoebe and the leading
hemisphere of Iapetus. Based on the wavelength
shifts of these bands, we find four separate 
types of trapped CO$_2$ on the satellites.
On the
inner satellites, CO$_2$ appears trapped in amorphous
ice sourced from Saturn's E-ring, and a second
component of CO$_2$ is associated with the dark
material most prominent on the trailing hemispheres
of Dione and Rhea. On the outer satellites,
CO$_2$ appears to be produced by irradiation of
organics on Phoebe, which are then transported to the 
dark leading hemisphere of Iapetus and onto the
dark regions of Hyperion. CO$_2$ is also trapped
by water ice on the trailing hemisphere
of Iapetus and on Hyperion. These observations 
point to the continued need for laboratory
studies to better understand 
the sources and trapping mechanisms of CO$_2$ 
throughout the outer solar system.

\end{abstract}

\section{Introduction}
\begin{figure*}
\begin{center}
\hspace*{-2cm}\includegraphics[scale=1]{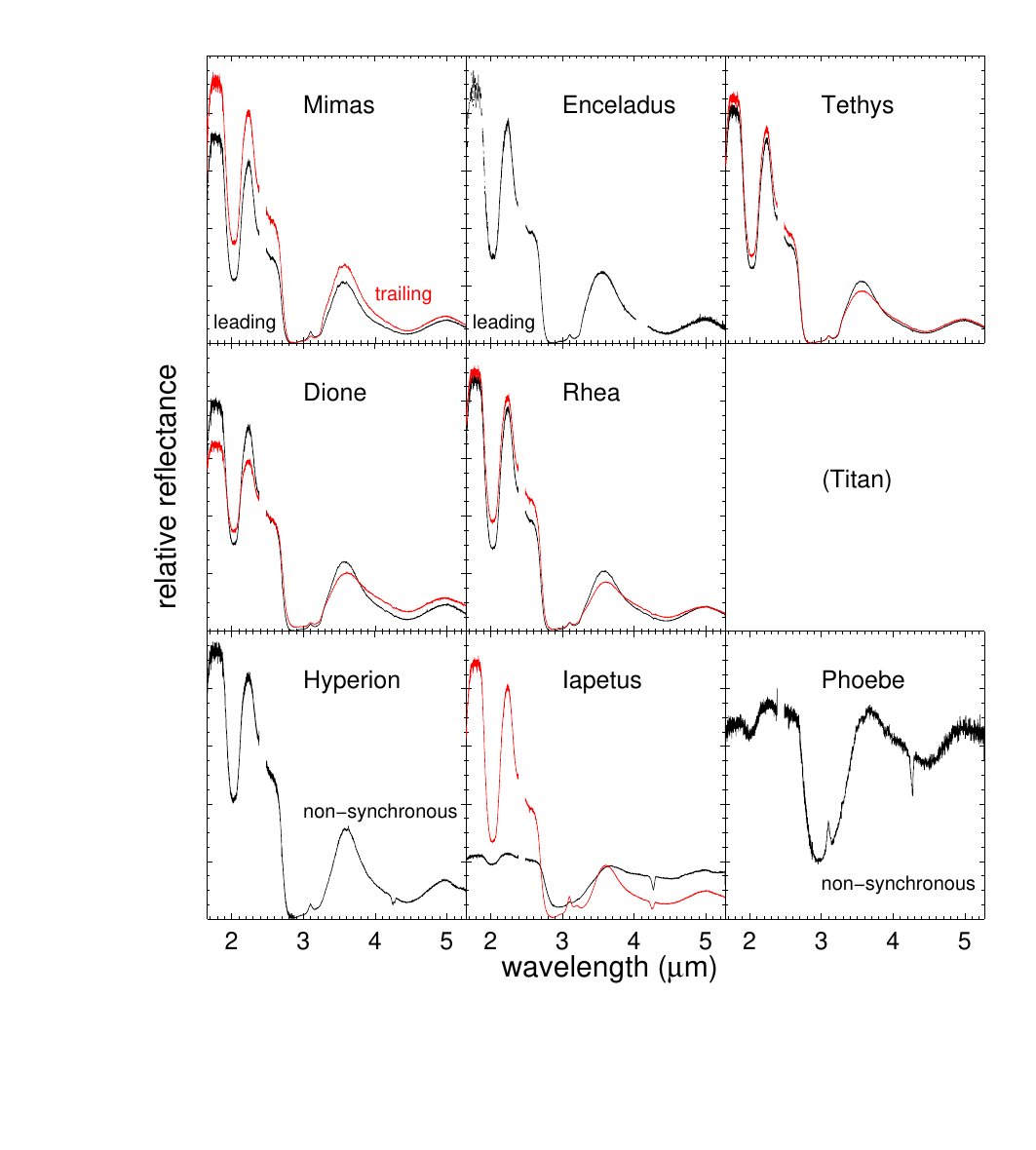}
    \caption{JWST spectra of the satellites of Saturn, in order
    of distance from Saturn. Titan's location is marked to 
    distinguish the inner and outer satellite systems. For synchronously
    rotating satellites, the leading hemisphere is shown in black 
    while the trailing is in red. For Enceladus only the leading
    was observed. For the non-synchronous Hyperion and Phoebe, only
    a single observation was made. The leading/trailing hemisphere spectra
    of individual satellites are scaled correctly relative to each other, 
    but the scaling of the spectra
    of the different satellites is arbitrary. { Uncertainties
    in the spectra can be estimated from the scatter in the data 
    of these generally smoothly varying spectra.}
    }
    \end{center}
\end{figure*}

Solid state CO$_2$ is seen throughout the outer solar system, even at temperatures where 
crystalline CO$_2$ should not be stable over the age of the solar system
\citep{ahrens_geoscientific_2022}. It is present on at least
one Jupiter Trojan asteroid \citep{Wong_JWST_2024} and on all three icy Galilean satellites \citep{mccord_non-water-ice_1998,trumbo_distribution_2023, villanueva_endogenous_2023,bockelee-morvan_composition_2024,cartwright_revealing_2024}, and
has been detected on most of the larger icy Saturnian satellites \citep{clark_compositional_2005,buratti_cassini_2005,brown_observations_2006,clark_compositional_2008,cruikshank_carbon_2010}.  CO$_2$ is
also seen on Centaurs well inside the frost stability line of
approximately 15 AU \citep{lebofsky_stability_1975}. The presence
of CO$_2$ above its stability temperature implies that CO$_2$ molecules are trapped within some other 
material. Suggestions have included trapping in crystalline or amorphous water ice \citep{cruikshank_carbon_2010, bockelee-morvan_composition_2024, schiltz_characterization_2024}, in organics \citep{villanueva_endogenous_2023, cartwright_revealing_2024},  
in mineral structures \citep{hibbitts_adsorption_2012}, and in other materials
\citep{ahrens_geoscientific_2022}. 
The source of the CO$_2$ is likewise unclear.
Possibilities suggested include that the CO$_2$ is native to the body \citep{hibbitts_carbon_2003}, that it is created from irradiated organics
or carbonates \citep{bockelee-morvan_composition_2024, cartwright_revealing_2024}, or that it might be a product of exogenous infall from comets or chondritic material
\citep{carlson_europas_2009}.
Neither the source of the CO$_2$ nor its trapping
mechanism has been firmly identified for any of these
bodies.

One powerful method for understanding the trapping mechanism 
(which might be distinct from the source mechanism) of CO$_2$ 
is the precise position
of the $\nu_3$ { asymmetric stretch} band and of the $\nu_1+\nu_3$ { combination band (C-O stretch + asymmetric stretch).}
In crystalline CO$_2$ at 80 K, these bands appear
at 4.268 $\mu$m and 2.698 $\mu$m, respectively \citep{oancea_laboratory_2012}.
When a CO$_2$ molecule is trapped within a foreign material, the molecular shape and bonds are
distorted slightly, yielding wavelength shifts of several nanometers or more. Each trapping mechanism
yields a characteristic shift, though in some cases different mechanisms can yield similar shifts,
rendering identification ambiguous. { Shifts in
the 4.268 $\mu$m and 2.698 $\mu$m bands are
independent, and combined can better distinguish
different trapping mechanisms.}

While CO$_2$ was originally detected from the Galileo spacecraft (for the Galilean satellites) and
the Cassini spacecraft (for the Saturnian satellites), recent observations of the icy Galilean satellites 
from JWST have shown that at higher sensitivity and higher spectral resolution, CO$_2$ on the surfaces
of these bodies is considerably more complex than 
previously understood. Europa has CO$_2$
most strongly concentrated in Tara Regio -- { a region
of extensive chaos} -- with a $\nu_3$ doublet at 4.249 and 4.268 $\mu$m
\citep{trumbo_distribution_2023, villanueva_endogenous_2023}. The longer wavelength is consistent with crystalline CO$_2$ which should be extremely unstable
at equatorial regions like Tara Regio. Interestingly, a 2.697 $\mu$m  singlet also appears, again at a similar
wavelength to crystalline CO$_2$. \citet{villanueva_endogenous_2023} model the CO$_2$ doublet as a combination of
trapping in methanol ice  and the presence of crystalline CO$_2$, while
\citet{trumbo_distribution_2023} point to lack of a detection of any
organic materials and the instability of crystalline CO$_2$ and
suggest trapping of oceanic CO$_2$ gasses or the irradiation of carbonates as possible sources and trapping
mechanisms. 

Ganymede shows a broad, variably asymmetric $\nu_3$ band with 
central wavelengths between 4.257 and 4.260 $\mu$m 
at low latitudes, where the wavelength is correlated with the optical
albedo and thus possibly related to the presence of water ice.
At high latitudes on the leading hemisphere the line shifts to 4.2695 $\mu$m, close
to the expected wavelength of crystalline CO$_2$
\citep{bockelee-morvan_composition_2024}. While the high latitude CO$_2$ 
is plausibly trapped in amorphous water ice, the source and trapping
mechanism of the low latitude CO$_2$ remains unknown. 
Callisto has the strongest CO$_2$ absorption of the Galilean satellites
and shows a 4.250 $\mu$m band on the trailing hemisphere
that shifts as far redward as 4.258 $\mu$m in the regions near
the Asgard basin \citep{cartwright_revealing_2024}. No satisfactory 
explanation of the source or trapping of the CO$_2$ has been made.

For the Galilean satellites, precise identification of sources and trapping mechanisms is stymied by the 
range in temperatures, compositions, irradiation environment, and subsurface structure of each of the satellites. 
The icy Saturnian satellites, in contrast, present a 
more uniform case study in trapped CO$_2$. 
The inner satellites (interior to Titan) have surfaces that are
nearly pure water ice, with only trace amounts of localized dark material, while the outer satellites (external to Titan) 
include more abundant dark material,
presumably sourced from Phoebe and its associated ring \citep{2018eims.book..307H}. 
The similarities and
differences in CO$_2$ abundances and wavelength shifts on these bodies
can allow us to disentangle the effects of ice vs. non-ice material,
the effects of organics, and the effects of the sources and trapping mechanisms
throughout the Saturnian system. While observations from Cassini/VIMS 
were capable of detecting the CO$_2$ on these objects \citep{cruikshank_carbon_2010}, observations with JWST allow 
both significantly higher sensitivity and higher spectral resolution. 
Here, we take advantage of these characteristics to study trapped CO$_2$
across the Saturn system.

\section{Observations}
We obtained JWST NIRSpec spectra from 1.8 to 5.2 $\mu$m of Mimas, Tethys, Dione, and Rhea -- mid-sized icy satellites interior to Titan, and of Hyperion, Iapetus, and Phoebe -- mid-sized satellites exterior to Titan
with at least some dark material. In all cases except Hyperion and Phoebe, which orbit non-synchronously,
we separately obtained spectra of the leading and of the trailing hemispheres. Observational details are given in Table 1. The observations were
obtained using the { G235H} grating at the shortest wavelengths, to prevent saturation, and the G395M grism
at the longer wavelengths, to increase the sensitivity to CO$_2$. In addition, we analyze data from
Program 1250 of { G235H} and G395H spectra of the leading hemisphere (only) of Enceladus. 
In our observations, all targets were observed with 4 dithers in the NIRSpec IFU (the Enceladus 
observations obtained only two dithers). The target was visually confirmed in the data, then the
data were processed through the empirical PSF-fitting pipeline described in \citet{2025PSJ.....6...22B} { using JWST CRDS context file jwst\_1252.pmap and pipeline version 1.14.0 \citep{2022zndo...7041998B}}.
Observations of the Solar analog P330E \citep{colina_absolute_1997}, from Program 1538, were reduced identically. The stellar 
spectrum was resampled at the wavelengths corresponding to the combined Doppler shift of the star, 
and the satellite geocentric and heliocentric velocity. The satellite spectrum was then divided by
the stellar spectrum to yield a reflectance spectrum with the solar lines removed, as shown in Figure 1. { While the JWST pipeline returns uncertainties, we find that empirically derived uncertainties determined
by the scatter of the data about a polynomial fit (for regions with
no absorption features) provides a more robust result, and such
uncertainties are used throughout.}

\begin{deluxetable*}{lccccccc}
\tablecaption{JWST observational parameters}
\tablehead{\colhead{satellite} & \colhead{date and time} & \colhead{exposure} & \colhead {$r$} & \colhead{$\dot r$} &\colhead{$\Delta$} & 
\colhead{$\dot\Delta$} & \colhead{longitude} \\
 & & \colhead{(s)} & \colhead{(AU)} & \colhead{(km s$^{-1}$)} &\colhead{(AU)} & \colhead{(km s$^{-1}$)} & \colhead{(deg)} }

\startdata
Mimas (leading)&       31-Oct-2023 05:47 &    291 & 9.75  &-12.3  &9.32  & 14.4  & 52\\
Mimas (trailing)     & 24-Jul-2024 11:03 &    291 & 9.67  &13.9  &8.95  & -6.5  & 262\\
Tethys (leading)     & 30-Oct-2023 18:39 &    116 & 9.75  &-11.2  &9.31  & 15.7  & 78\\
Tethys (trailing)    & 20-Jul-2024 08:03 &    116 & 9.67  &10.7  &9.00  & -11.1  & 271\\
Dione (leading)      & 09-Nov-2023 01:38 &    116 & 9.75  &-10.4  &9.46  & 18.3  & 98\\
Dione (trailing)     & 22-Jul-2024 01:36 &    116 & 9.67  &9.37  &8.98  & -11.9  & 274\\
Rhea (leading)       & 27-Nov-2023 05:27 &    116 & 9.74  &-8.30  &9.76  & 20.7  & 76\\
Rhea (trailing)      & 12-Jul-2024 13:24 &    116 & 9.68  &7.88  &9.10  & -16.3  & 261\\
Hyperion      &        29-Oct-2023 05:57 &    350 & 9.75  &-4.9  &9.29  & 21.7  & - \\
Iapetus (leading)    & 16-Oct-2023 17:28 &    116 & 9.76  &-3.6  &9.11  & 19.8  & 80\\
Iapetus (trailing)   & 25-Jul-2024 15:56 &    116 & 9.67  &2.6  &8.93  & -17.5  & 267\\
Phoebe               & 20-Nov-2023 22:50 &    1050 & 9.77  &-2.4  &9.69  & 26.9  & 65\\
\enddata
\tablecomments{Exposure times are identical for the
G235H and G395M settings. $r$ and $\Delta$ refer
to the heliocentric and JWST-centric distances,
respectively, with $\dot r$ and $\dot \Delta$
the heliocentric and JWST-centric velocities.
Longitude is the sub-observer longitude at the time
of observation. Hyperion has no defined longitude
system.}
    
\end{deluxetable*}

At the broad scale discernible in Figure 1, the main features visible are the distinct absorptions
due to water ice, with bands at 2, 3 and 4.5 $\mu$m. For the high-albedo nearly-pure water
ice surfaces, the 3 $\mu$m absorption feature goes to essentially zero reflectivity at 2.87~$\mu$m. Phoebe and the
leading hemisphere of Iapetus have the least coverage of water ice, as can be seen in their
elevated 3 $\mu$m reflectivities, and Hyperion and the trailing hemispheres of Dione and Rhea also have
slightly non-zero 3 $\mu$m reflectivities. Also visible is the 3.1 $\mu$m Fresnel reflection peak. { The wavelength of this peak shows the
presence of ice in crystalline form, though the 3.2 $\mu$m secondary
peak expected for pure crystalline ice is subdued (except on the trailing
hemisphere of Iapetus) allowing the possibility of at least some
amorphous ice in the top layers}
\citep{mastrapa_optical_2009}. 
{ Finally, slight shifts in the $\sim$3.6 $\mu$m}
peak between the 3 and 4 $\mu$m bands show temperature differences, with the trailing hemispheres of
Dione and Rhea shifted to slightly longer wavelengths, indicative of higher temperatures, as would
be expected for these slightly darker surfaces. 
\bigskip
\bigskip
\bigskip

\section{Results}
\begin{figure*}
\begin{center}
\hspace*{-2cm}\includegraphics[scale=1]{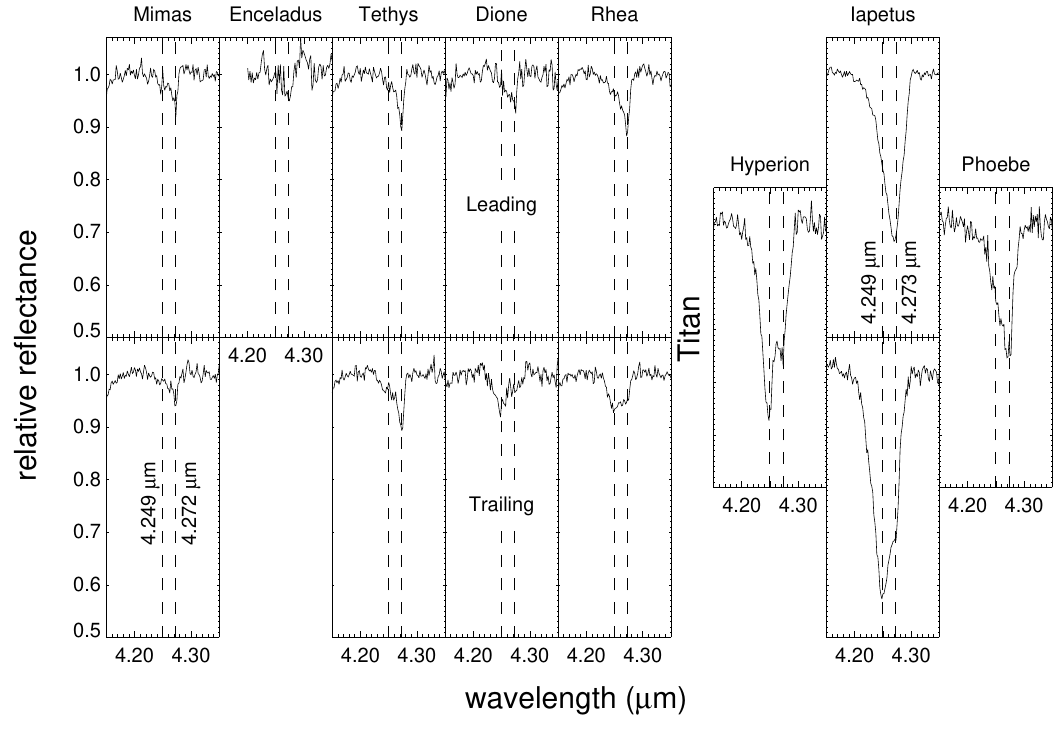}
    \caption{The region near the 4.27 $\mu$m $\nu_3$ band of CO$_2$ for the Saturnian
    satellites. { The top row shows the leading hemispheres,
    while the bottom shows the trailing. For non-synchronous Hyperion
    and Phoebe, only a single spectrum is shown. Each spectrum was divided by a local continuum to examine relative
    absorption, and the uncertainties can be estimated by examining the
    scatter of the data about the continuum.} All satellites have absorptions near 4.25 and 4.27 $\mu$m or both,
    though the central wavelengths of the absorption on the outer satellites differ
    slightly from those of the inner satellites. For the inner satellites, dashed lines
    are shown at 4.249 and 4.272 $\mu$m, while for the outer satellites dashed lines
    are shown at 4.249 and 4.273 $\mu$m.}
    \end{center}
\end{figure*}
An absorption band near the 4.27 $\mu$m $\nu_3$ CO$_2$ region is detected on every hemisphere of every satellite observed, even where the band was not detectable in Cassini data. 
To more closely examine 
this spectral region, we divide each spectrum by an estimated continuum, obtained
by performing a second-order polynomial fit to the region between 4.15 and 4.40 $\mu$m,
excluding the region between 4.20 and 4.30 $\mu$m where CO$_2$ absorption
might be present. We perform the same fit on data from the leading hemisphere
of Enceladus, taken at higher spectral resolution in Program 1250. For the Enceladus
spectrum we rebin by a factor of 3 to approximate the same spectral resolution
as the other satellites. The results are shown in Figure 2. All
observations show absorptions near 4.25 $\mu$m, 4.27 $\mu$m,
or both.
The absorptions on the outer satellites (defined by those beyond the orbit of Titan) are significantly stronger than those on the inner
satellites and at slightly different wavelengths.
{ We perform a two-Gaussian fit to each of the continuum-divided
spectra in the  4.27 $\mu$m region to obtain best fit models 
and uncertainties (where, again, the uncertainties in the data
are estimated as the scatter in the region outside of the 
absorption features). The results of the fits are
given in Table 2.}
\begin{rotatetable*}
\begin{deluxetable*}{|ll|lll|lll|lll|}
\tablecaption{CO$_2$ band measurements}
\tablehead{\colhead{satellite} & \colhead{~ } & \colhead{wavelength} & \colhead{width} & \colhead{strength} & \colhead{wavelength} & \colhead{width} & \colhead{strength} & \colhead{wavelength} & \colhead{width} & \colhead{strength} \\
\colhead{~} & \colhead{~} & \colhead{($\mu$m)} & \colhead{($\mu$m)} & \colhead{} &  \colhead{($\mu$m)} & \colhead{($\mu$m)} & \colhead{} &\colhead{($\mu$m)} & \colhead{($\mu$m)} & \colhead{}}

\startdata
Mimas& leading & 2.705(1) & 0.010(1)&  0.043(4)   & 4.25(2) & 0.017(2) & 0.034(3) & 4.2715(5)  & 0.0040(6)  & 0.059(7) \\
Mimas& trailing &2.7082(4)& 0.0062(4)& 0.080(4)   & 4.256(5)& 0.017(4) & 0.029(6) & 4.273(7)   & 0.003(1)   & 0.05(2)  \\
Enceladus&      & 2.7074(3)&0.0060(3)& 0.063(3)   & \nodata & \nodata  & $<0.02$  & 4.272(2)   & 0.006(2)   & 0.06(1)  \\
Tethys& leading &2.7065(4)& 0.0066(4)& 0.069(4)   & 4.256(3)& 0.017(2) & 0.037(3) & 4.2720(4)  & 0.0041(5)  & 0.085(8) \\
Tethys& trailing&2.7062(6)& 0.0095(6)& 0.0095(6)  & 4.255(4)& 0.016(3) & 0.045(6) & 4.2724(6)  & 0.0042(8)  & 0.09(1)  \\
Dione& leading & 2.7081(3) & 0.0056(3) & 0.081(3) & 4.262(2)& 0.011(2) & 0.048(5) & 4.2738(7)  & 0.0022(7)  & 0.05(1)  \\
Dione& trailing &2.7086(6) & 0.0081(6) & 0.037(2) & 4.248(1)& 0.012(1) & 0.072(5) & 4.277(5)   & 0.008(2)   & 0.032(6) \\
 Rhea& leading & 2.7079(4) & 0.0073(3) & 0.065(3) & 4.252(2)& 0.020(1) & 0.057(3) & 4.2712(4)  & 0.0051(5)  & 0.085(7) \\
 Rhea& trailing &2.7092(3) & 0.0057(3) & 0.069(3) & 4.252(2)& 0.014(2) & 0.076(4) & 4.272(1)   & 0.005(2)   & 0.03 (1)\\
Hyperion     & & 2.6975(1) & 0.0007(1) & 0.16(3)  & 4.2473(7)&0.0147(6)& 0.347(6) & 4.2758(8)  & 0.0079(8)  & 0.18(1)  \\
Iapetus& leading & \nodata & \nodata   &$<0.02$   & 4.251(3)& 0.018(1) & 0.13(2)  & 4.2724(4)  & 0.0124(4)  & 0.24(3)  \\
Iapetus& trailing &2.706(1)& 0.015(1)  &0.033(3)  & 4.2507(3)&0.0189(2)& 0.414(4) & 4.2739(3)  & 0.0043(4)  & 0.113(9) \\
Phoebe        & & \nodata  & \nodata   &$<0.03$   & 4.260(1)& 0.0163(7)& 0.16(1)  & 4.2735(5)  & 0.0060(7)  & 0.13(2)  \\
   \enddata

   \tablecomments{The wavelength, width, and strength of the absorption are determined by one (for the 2.7$ \mu$m region) or two (for the 4.27 $\mu$m region) gaussian fits to the data. The uncertainty in the last significant digit is given in parenthesis. The strength is the peak of the gaussian fit to the continuum-removed data, while the
   width is the $1\sigma$ gaussian width. For non-detections, a 1 $\sigma$ upper limit to the absorption depth is given.}
   \end{deluxetable*}
\end{rotatetable*}

In addition to the $\sim$4.27 $\mu$m $\nu_3$ absorption, Hyperion and the 
trailing hemisphere of Iapetus have narrow absorptions near the location
of the $2.7 \mu$m $\nu_1+\nu_3$ combination band. Many of the other satellites appear
to have a broader feature at slightly longer wavelength. 
To examine these features more 
closely, 
we perform a similar analysis as above to the data, now performing a third-order
polynomial fit to the region between 2.660 and 2.760 $\mu$m, excluding
the regions between 2.690 and 2.703 and between 2.705 and 2.717 $\mu$m.
Results are shown in Figure 3.

From the continuum-divided spectra, 
{ narrow absorptions on Hyperion and the trailing hemisphere of Iapetus are evident as well as
a broad absorption centered around 2.71 $\mu$m on many of the
satellites. Uniquely, Phoebe and the dark leading hemisphere of Iapetus show
no clear signs of absorption near 2.7 $\mu$m, while only Hyperion and
the bright trailing hemisphere of Iapetus show the narrow
band near 2.70 $\mu$m. We determine precise wavelengths, strengths, and widths of these absorption
features by fitting the continuum-removed data to a single gaussian. The
results are given in Table 2.}

The combination of the spectral characteristics near 4.27 $\mu$m 
and near 2.7 $\mu$m makes a crucial distinction clear:
while the inner and outer satellites share similar sets of absorptions
in the 4.27 $\mu$m region, the differences in the 2.7 $\mu$m 
region -- { including the narrow 2.70 $\mu$m absorptions on Hyperion and the trailing hemisphere of Iapetus and the lack
of absorption in the 2.7 $\mu$m on Iapetus and Phoebe} -- require that the trapped state of the CO$_2$ on the inner and outer
satellites is different, regardless of the similarity of the wavelengths of the
4.27 $\mu$m absorptions. We thus consider the inner and outer satellites
separately below.

\begin{figure*}
\begin{center}
\hspace*{-2cm}\includegraphics[scale=1]{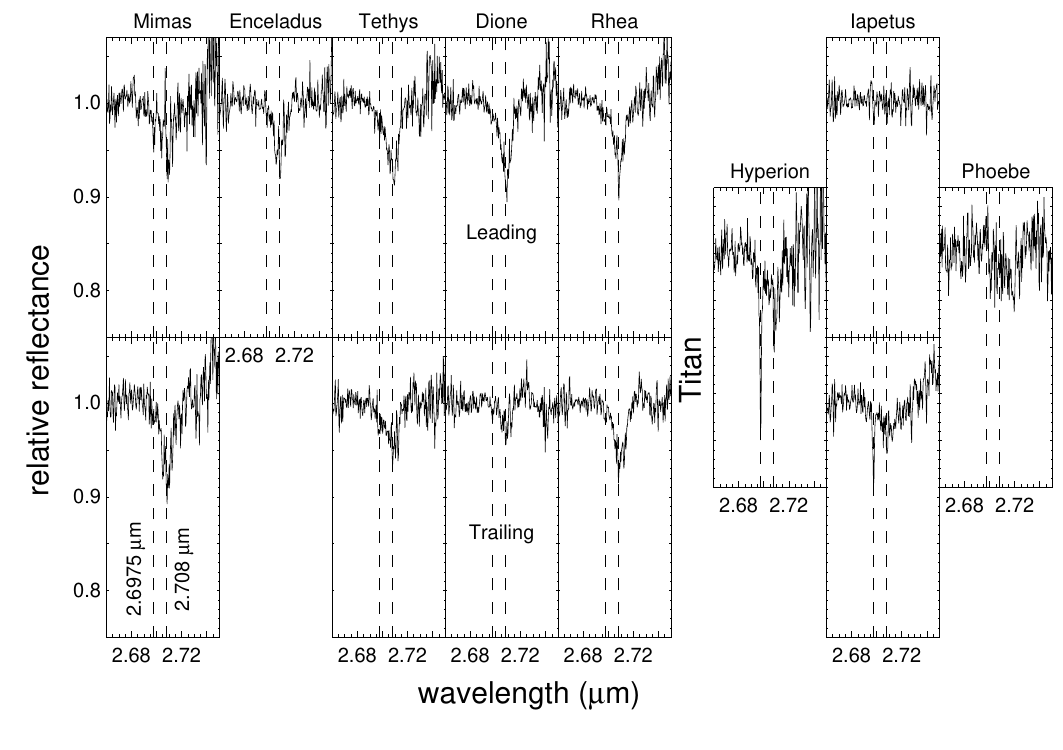}
    \caption{Continuum-divided spectra of the 2.7 $\mu$m region near 
    the $\nu_1+\nu_3$ combination band of CO$_2$. Most satellites have
    a broad absorption near 2.71 $\mu$m, while Hyperion and the bright trailing
    hemisphere of Iapetus have a narrow absorption near 2.70 $\mu$m. Phoebe and the
    dark leading hemisphere of Iapetus have no discernible absorption. All spectra 
    are shown with dashed lines at 2.6975 and 2.708 $\mu$m for clarity.}
    \end{center}
\end{figure*}

\subsection{The inner icy satellites}
The CO$_2$ signatures of the mid-sized icy satellites interior 
to Titan have strong similarities. In all cases, an { absorption
near
4.272$\mu$m
is present, while on the trailing hemispheres of Dione and Rhea,
a feature near 4.250 $\mu$m also appears and the feature near 4.272 $\mu$m appears 
diminished.  Likewise, a broad  feature near 2.708 $\mu$m is present on the inner
satellites,
with the weakest such feature seen on 
Dione's trailing hemisphere. 

Based on the spectral similarities of the 
satellites, we assume that they each have the same sets of spectral features. We can thus obtain a higher
signal-to-noise spectrum -- allowing a more detailed view
of the absorptions -- by taking spectral averages.
We first average 
both hemispheres of Mimas and Tethys 
and the leading hemispheres of Dione and Rhea to represent the spectra
of surfaces dominated by the feature near 4.272 $\mu$m.
Second we average 
the trailing hemispheres of Dione and Rhea to represent the surfaces
more dominated by the feature near 4.250 $\mu$m. 
These average spectra are
shown in Figure 4.
Performing gaussian fits identical to those done for the individual spectra, we find that
in both averaged spectra the short wavelength bands are centered at 2.708$\pm$0.002$\mu$m,
while the strongest longer wavelength band is found at 4.2724$\pm$0.0002 $\mu$m for the first
average and 4.249$\pm$0.001 $\mu$m for the second set of averaged spectra. }
\begin{figure*}
\begin{center}
\hspace*{-2cm}\includegraphics[scale=1]{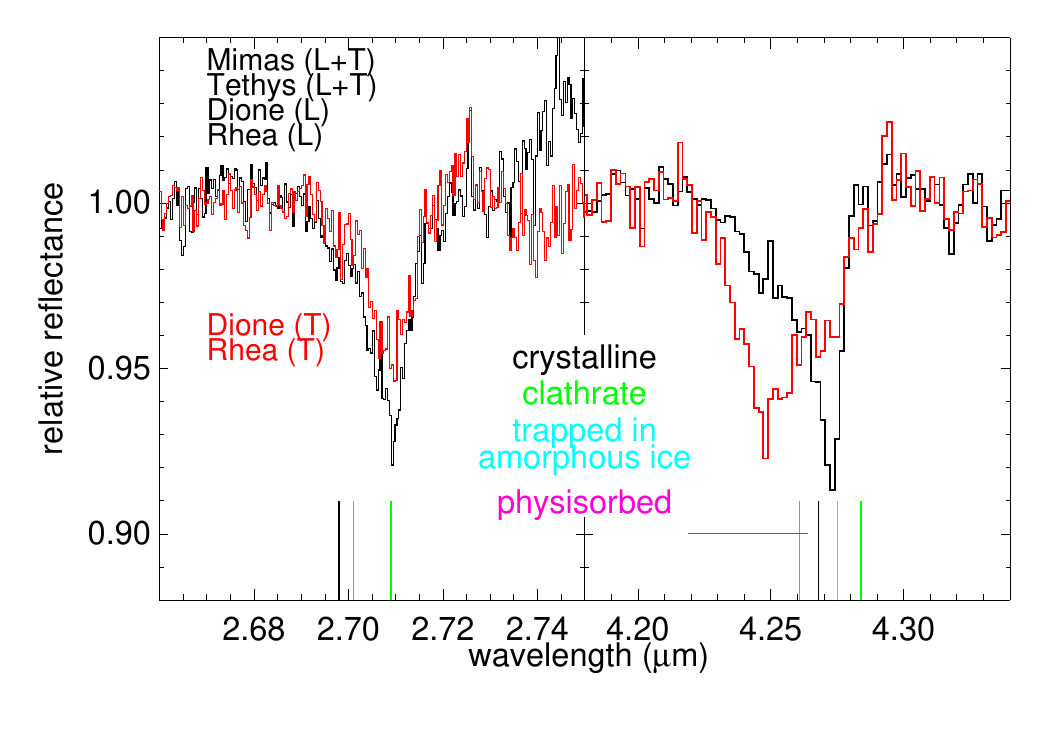}
    \caption{An average of the continuum-removed spectra of the leading
    and trailing hemispheres of Mimas and Tethys and the leading hemispheres
    of Dione and Rhea, in black, and of the trailing hemispheres of Dione and 
    Rhea, in red. The lines at the bottom show measured wavelengths of
    CO$_2$ in different states, including in pure crystalline form and trapped
    as clathrate in crystalline water ice \citep{oancea_laboratory_2012},
    trapped in amorphous ice \citep{sandford_physical_1990, galvez_trapping_2008}  and 
    physisorbed onto minerals \citep{hibbitts_physisorption_2007}, which can lead to a wide range of wavelengths. The shorter wavelength 4.2487 $\mu$m
    line associated with the darker hemispheres of Dione and Rhea appears to have 
    no corresponding absorption in the 2.7 $\mu$m region.
    }
    \end{center}
\end{figure*}

Based both on optical imaging \citep{schenk_plasma_2011} and the depth of the 3 $\mu$m  absorption, 
the trailing hemispheres of Dione and Rhea have the strongest
concentrations of dark non-water ice material. The presence of
the shorter wavelength 4.249 $\mu$m  absorption on the trailing sides of these
two satellites { leads to the hypothesis} that this absorption is associated with
the trapping of CO$_2$ in this dark material and that the
weakness of the 4.272 $\mu$m  absorption on these trailing hemispheres
 is due to this extra covering of dark material.
{ While this shorter wavelength feature is strongest on these two
 satellites, it appears to be present on
 all of the inner satellite spectra,} where
the asymmetric 
nature of the 4.273 $\mu$m band, with a broad wing to shorter wavelengths,
is plausibly caused by the blending of a small amount of
the shorter wavelength absorption. Indeed, optical imaging shows
that the dark material is
seen throughout the inner satellites, even if dominant on the trailing
hemispheres of Dione and Rhea \citep{schenk_plasma_2011}.

The composition of the dark material on the inner satellites remains unidentified,
but the shift of the $\nu_3$ line to shorter wavelength
can be caused by the trapping of the CO$_2$ molecule into
a tightly confined space \citep{sandford_physical_1990}.
For example, cryogenic experiments
with trapping in zeolites and clays showed absorption
features ranging from 4.219 to 4.264 $\mu$m  \citep{hibbitts_physisorption_2007}.
Clark et al. (\citeyear{clark_compositional_2005, clark_compositional_2008}) have argued that the dark material throughout
the Saturnian system is a fine-grained material, possibly a nanophase hematite
or iron. While no relevant spectroscopic studies of CO$_2$ trapped by
these types of materials or studies at these temperatures
have been performed, such small iron oxides could be
efficient at CO$_2$ capture \citep{baltrusaitis_carbon_2011, hakim_studies_2016}.

The association of the shorter wavelength 4.249 $\mu$m absorption
with the dark material suggests that the { longer wavelength feature and the 2.708 $\mu$m
absorption
could be due
to CO$_2$ trapped in the water ice which dominates the remainder of these
surfaces.
Indeed, the measured wavelength for both spectral 
lines (2.708$\pm$0.002 and 4.2724$\pm$0.0002 $\mu$m) is nearly coincident with the 
values measured (2.705 and  4.2737 $\mu$m) for
 CO$_2$ absorption
in a 100 K co-deposited 50:1 mixture of amorphous water ice and CO$_2$   \citep{sandford_physical_1990}.}

The inner satellites are all embedded within Saturn's E-ring which is
sourced from the geysers on the south
pole of Enceladus \citep{spahn_cassini_2006}.
These particles themselves are 
composed of water ice with possible
inclusions of silicates, organics, 
and CO$_2$ \citep{hillier_composition_2007,postberg_e-ring_2008}.
{ While observations of satellite albedo, color, infrared spectra, and particulate grain sizes
have led to varying conclusions about the interaction between the surfaces of
the satellites and the E-ring particles \citep[i.e.][] {2018eims.book..343H}},
Cassini radar observations suggest
that these inner satellites are
coated with decimeters of fresh clean
water ice from the E-ring
\citep{le_gall_dust_2019}.
{ If correct, it is likely that the observed
CO$_2$ is trapped in amorphous water carried in from the E-ring particles.
The actual source of the CO$_2$ is less certain. While it could plausibly be
carried in with the E-ring particles { or formed from radiolysis
of embedded organic materials}, it is also possible that the CO$_2$
is native to each individual satellite and is released from the surface or subsurface
only to be trapped by the ice.}

\subsection{The outer satellites}
\begin{figure*}
\begin{center}
\hspace*{-2cm}\includegraphics[scale=1]{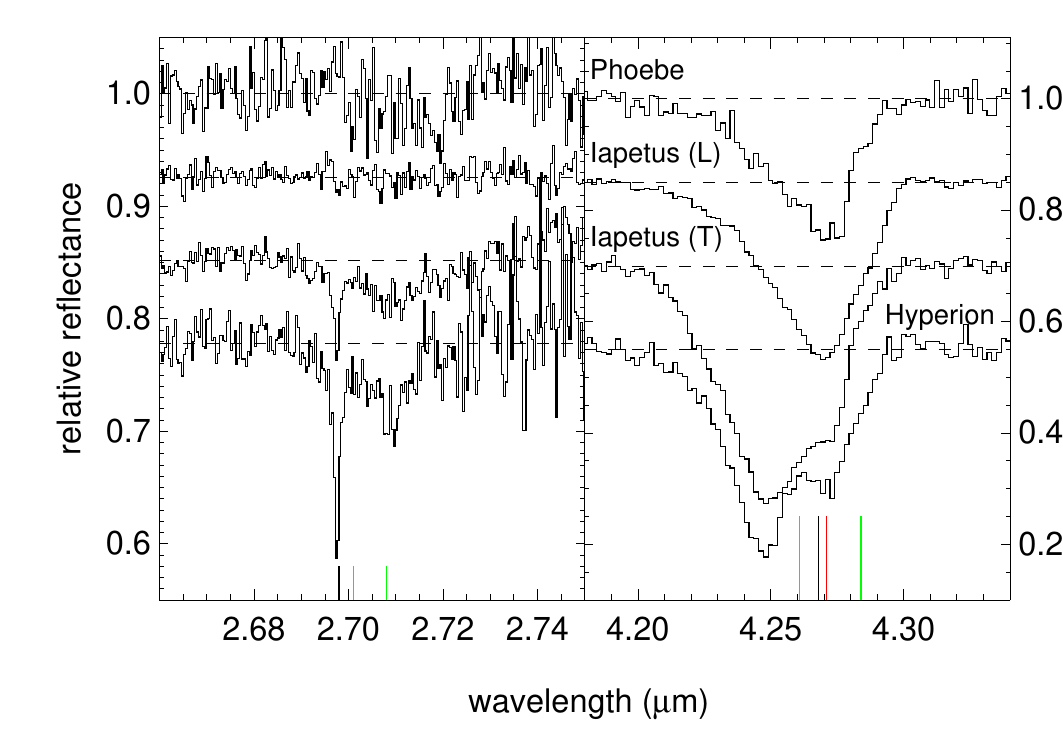}
    \caption{Continuum-removed views of the 2.7 and 4.27 $\mu$m regions of the spectra
    of the outer satellites. The colored lines show the wavelengths of various forms
    of CO$_2$, as in the previous Figure. The black line again shows the wavelength of crystalline
    CO$_2$ and the green lines show those of CO$_2$ clathrates.
    The red line (not shown in the
    previous Figure), which is consistent
    with the wavelength seen in the high signal-to-noise Iapetus data, is the wavelength
    of CO$_2$ formed in irradiated organic material.
    }
    \end{center}
\end{figure*}

The three observed satellites that are exterior to Titan have stronger
CO$_2$ absorptions and appear to have different forms
of CO$_2$ than those interior to Titan. Figure 5 shows the continuum-removed spectra
from the 2.7 $\mu$m $\nu_1+\nu_3$ CO$_2$ absorption region and the 4.27 $\mu$m $\nu_3$
absorption region for these satellites. 
As with the inner satellites, there appear to be 
two separate trapped states of CO$_2$, reflected 
in a shorter and longer wavelength component of
the $\nu_3$ absorption. { The two lines both contribute strongly
and overlap across the band, so precise estimation of band centers is less
certain, but the shorter wavelength feature averages 4.2487$\pm$0.0004$\mu$m on the two 
spectra where the lines are strongest. Likewise, the longer wavelength feature
averages 4.2730$\pm$0.0003 $\mu$m on the two spectra where this component is strongest.}

On Phoebe, spatially resolved observations from Cassini found that the CO$_2$ absorption is most strongly associated with the dark material \citep{clark_compositional_2005}. 
The leading hemisphere of Iapetus
is known to be coated
with this dark material, which spirals inward through the large Phoebe ring 
\citep{verbiscer_saturns_2009}. The { location of the dominant}
CO$_2$ absorption on the leading hemisphere
of Iapetus matches that of Phoebe { to the 2$\sigma$ level.
The 4.2730$\pm$0.0003 average position of this line 
is close to, but
not coincident with, the 4.2724$\pm$0.0002 line of the inner satellites, though the strong blending of the lines
for the outer satellites makes the reliability of the differences between these values
less certain. }

We thus associate this 4.2730 $\mu$m  absorption
with the dark material on Phoebe and Iapetus.
The brighter trailing hemisphere of Iapetus and
the interiors of the craters on Hyperion
contain minor amounts of the dark material,
and indeed,
{ a long wavelength absorption appears weakly in spectra of
both the leading hemisphere of Iapetus and of Hyperion,
where they are blended with the stronger}, shorter wavelength absorption at 4.2487 $\mu$m. While the lower
signal-to-noise Phoebe data show some variability around 2.7 $\mu$m, the higher
signal-to-noise data on the dark hemisphere
of Iapetus makes it clear that this
dark material shows no 
clear indications of any absorption due
to the 2.7 $\mu$m $\nu_1+\nu_3$ band.

A plausible source for CO$_2$ in the dark material of Phoebe is the irradiation of
organic material in close proximity to water ice. Laboratory experiments have found
that irradiation of ice-covered organics at temperatures similar to that in the 
Saturnian system produce a $\nu_3$ CO$_2$ band at $4.271 \pm 0.002$ $\mu$m,
consistent with the wavelength measured here 
\citep{gomis_co2_2005}. Unfortunately,
these experiments do not examine the 2.7 $\mu$m
region so it is unclear if the 
lack of $\nu_1+\nu_3$ absorption seen on
Phoebe and the dark hemisphere of Iapetus
is also reproduced.
Interestingly, the C-H stretch absorption features in the 3.4 $\mu$m
region suggested to be
present by \citet{cruikshank_aromatic_2014} do not appear in these JWST data, though 
smaller irregular satellites do appear to have features in this region \citep{2025PSJ.....6...97B}. More analysis is required to understand the presence and state of organic 
material on these satellites, but the existence of at least some organic materials on a dark outer 
solar system body such as Phoebe seems likely.

The association of the longer wavelength 4.2730 $\mu$m absorption with the dark material
on the outer satellites suggests that either the Phoebe-sourced dark material and its associated CO$_2$
are transported to these other satellites or that the dark material is transported without
the CO$_2$ (or loses its CO$_2$ upon impact) and the CO$_2$ is then regenerated on the satellites themselves { through 
subsequent irradiation}. Some evidence for the latter hypothesis
is that spatially resolved Cassini spectra of these satellites suggest that the CO$_2$ absorption is strongest
in the transition regions between the bright and dark material, where the dark material
would have the highest access to water { which can aid the radiolytic 
production of CO$_2$} \citep{palmer_production_2011}.

If CO$_2$ is being continuously created on these satellites from the irradiation of freshly
delivered organic material, the strong CO$_2$ signature on the icy parts of these bodies
are then plausibly produced from the same mechanism but then separately trapped in that icy material. It is
surprising that CO$_2$ trapped in ice has a different wavelength in the outer satellite 
system than in the inner, but the presence of the 2.6975 $\mu$m absorption exclusively on
the brighter, icier trailing hemisphere of Iapetus and on water-ice rich Hyperion { suggests} that
the CO$_2$ is associated with ice on these satellites, but in a physically different way than on the inner satellites.

While some experiments on trapping of CO$_2$
in water ice have shown small secondary
absorption features near the 4.248 $\mu$m
feature seen here \citep{gudipati_thermal_2023,schiltz_characterization_2024},
no trapping mechanism in either crystalline or amorphous water ice has been 
measured to have a prominent peak at
this position. Additionally, no 
trapping mechanism has ever been found to make a narrow 2.6975 $\mu$m feature like those
seen here (though this weaker combination band is often not studied in laboratory experiments). These observations point to the need for substantially more laboratory work
on the trapping of CO$_2$ and point to plausible materials on which to focus future
work.

\subsection{$^{13}$CO$_2$}
{ With the detections of the strong CO$_2$ bands on the outer satellites, we consider the 
possibility of the detection of the 4.384 $\mu$m absorption due to
the isotopologue $^{13}$CO$_2$.
Fig. 6 shows the continuum-divided region near the predicted $^{13}$CO$_2$ line and
gaussian fits to the Hyperion, Iapetus (leading and trailing) and Phoebe spectra.
No absorptions are seen in most of the spectra, but the trailing hemisphere of Iapetus
shows a 4.3$\pm$0.5\% absorption centered at 4.383$\pm$0.001 $\mu$m, consistent
with the expected $^{13}$CO/$^{12}$CO$_2$ band. While ratios of band strengths between the $^{13}$CO$_2$ and $^{12}$CO$_2$ bands 
cannot be used to
directly infer the $^{13}$C/$^{12}$C ratio, as the trapping mechanisms can affect
these molecules differently, the values are informative for comparison between
the bodies.
We find $1\sigma$ upper limits 
of 0.020, 0.008, and 0.012 for Hyperion, the leading hemisphere of Iapetus, and
Phoebe, respectively, and we measure a ratio of $0.019\pm0.02$ for the trailing hemisphere of Iapetus. \citet{clark_isotopic_2019} report values 
measured from VIMS data on Iapetus (dark material) and Phoebe (full disk) of 
0.012$\pm$0.001 and 0.053$\pm$0.006, respectively. 
While the VIMS Iapetus value is a plausible combination of 
some fraction of the leading and trailing hemisphere values measured here, the 
high abundance of $^{13}$CO$_2$ reported for Phoebe would require a 4.9\% absorption
feature,  which  can be ruled out here at the 5$\sigma$ level. 

The $^{13}$CO$_2$/$^{12}$CO$_2$ ratio
measured for the trailing hemisphere of Iapetus is higher
than the upper limits measured for both Phoebe and for the leading 
hemisphere of Iapetus, reinforcing the conclusion above that
the Iapetus leading hemisphere CO$_2$ differs in source, trapping
mechanism, or both from
that of the trailing hemisphere and of Phoebe. 
The less stringent Hyperion upper limit
remains consistent with the measured value for the leading hemisphere of
Iapetus, in agreement with our suggestion that the CO$_2$
at these locations are related.
While we cannot directly infer the $^{13}$C/$^{12}$C ratio from
these results, we note that the ratios are broadly consistent with the measured interstellar
isotopic ratio of 0.014$\pm$0.003 \citep{2000A&A...353..349B} and with the terrestrial value of 0.011.}
\begin{figure*}
\begin{center}
\hspace*{-2cm}\includegraphics[scale=1]{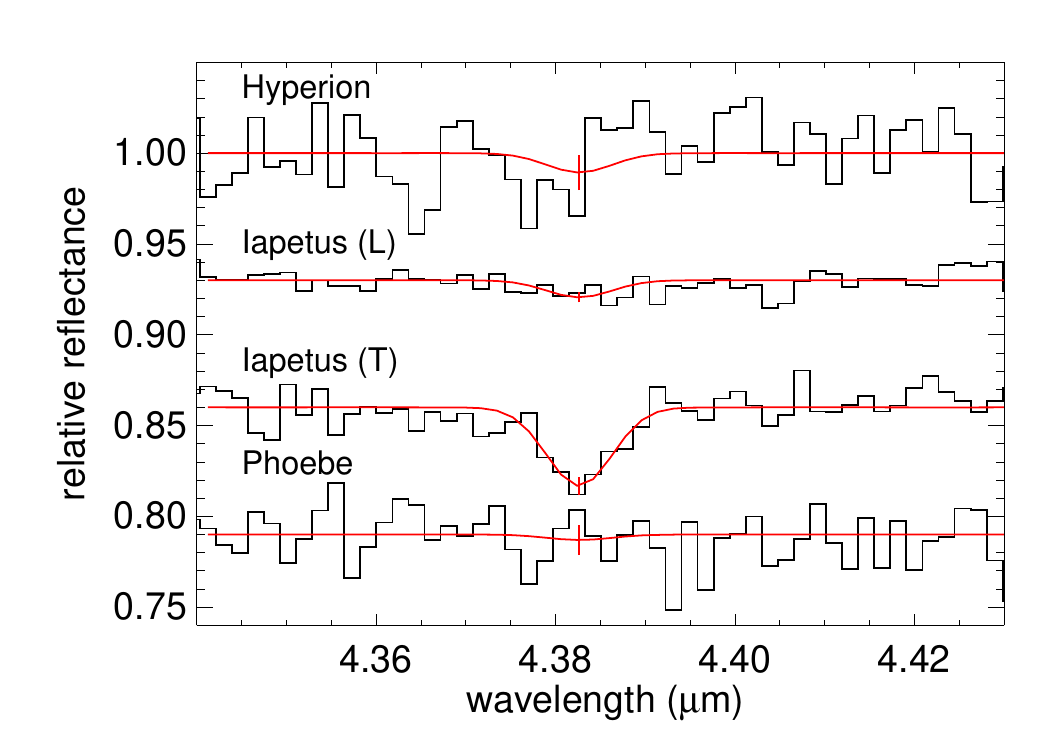}
    \caption{Continuum-divided region of the spectra containing the
    expected $^{13}$CO$_2$ absorption feature. Each spectrum is offset 
    by 0.07 units for clarity. A single gaussian fit to the data is
    used to determine detection limits or the strength of the absorption.
    The vertical red bar shows the 1$\sigma$ uncertainty on the line
    strength.}
    \end{center}
\end{figure*}

\section{Discussion} 
The satellites of Saturn have at least two separate sources of CO$_2$ and at least four separate
trapping mechanisms. 
{ The inner satellites have a CO$_2$ component with a $\nu_3$ absorption
at 4.2724$\pm$0.0002 $\mu$m and a broad $\nu_1+\nu_3$ combination band at 2.708$\pm$0.002 $\mu$m.
This CO$_2$ appears to be associated with the ice component of these satellites, with the
likely source either the E-ring particles or CO$_2$ native to each satellite.
A second component on the inner satellites has a $\nu_3$ absorption at 4.249$\pm$0.001 $\mu$m and no corresponding
$\nu_1+\nu_3$ absorption. This CO$_2$ appears to be trapped in the dark materials most
strongly seen on the trailing hemispheres of Dione and Rhea. The source could again be
native or could be the dark material itself.
The outer satellites have a CO$_2$ component with a $\nu_3$ absorption
at 4.2730$\pm$0.0003 $\mu$m that has no measurable $\nu_1+\nu_3$ component.
This CO$_2$ appears to be associated with the abundant dark material present on
Phoebe, the leading hemisphere of Iapetus, and parts of Hyperion. The CO$_2$ is
plausibly created from irradiation of organics associated with this dark material. 
A second CO$_2$ component on the outer satellites has a $\nu_3$ absorption
at 4.2487$\pm$0.0004 $\mu$m and a narrow $\nu_1+\nu_3$ band at 2.6975$\pm$0.0001 $\mu$m
as well as the same broad 2.708$\pm$0.002 $\mu$m band seen in the inner satellites.
This component appears associated with the icy regions of the outer satellites, and
the source could either be native or CO$_2$ released from the darker regions.}
\begin{figure}
    \includegraphics[scale=.5]{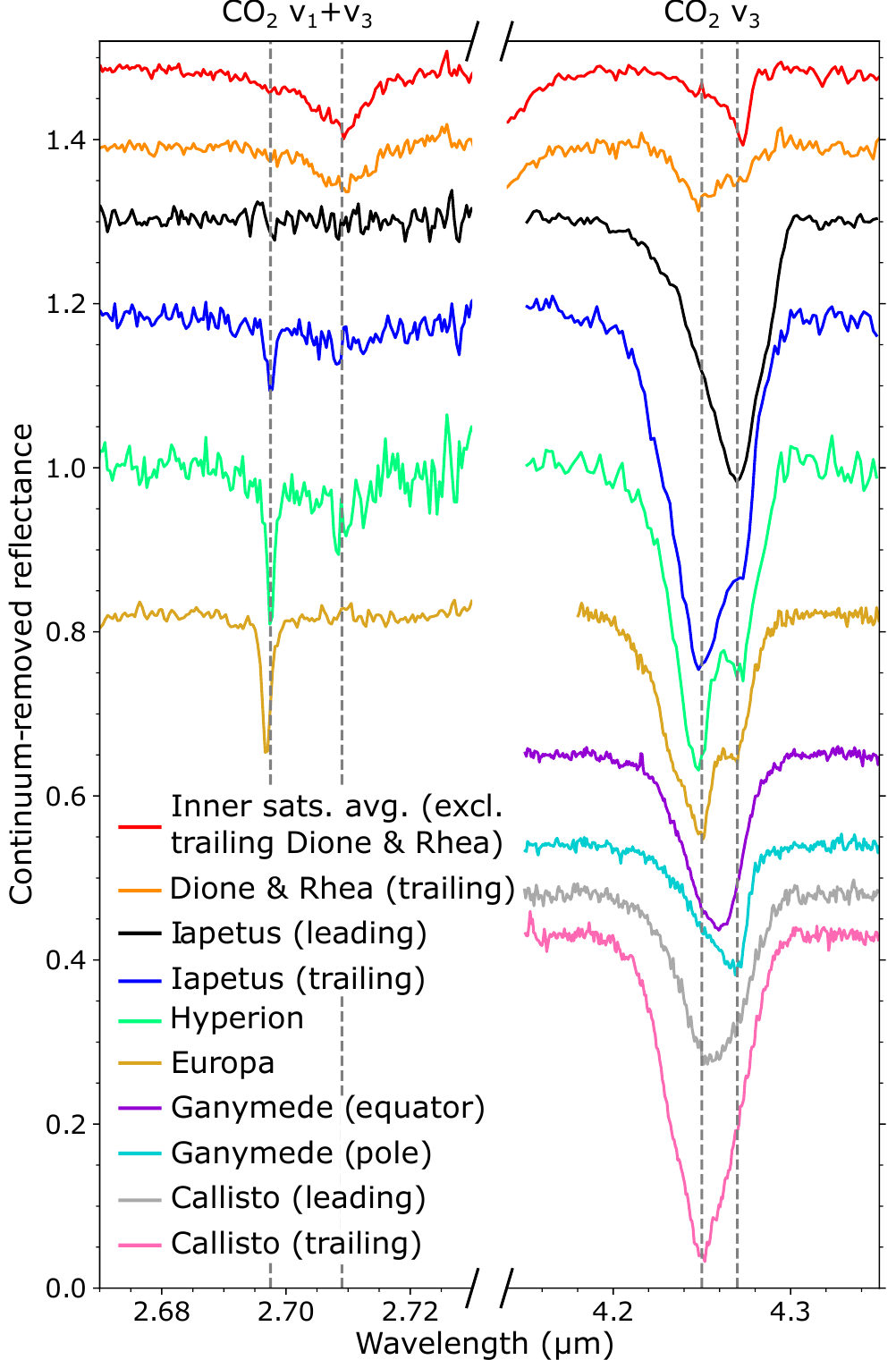}
    \caption{
    A comparison of CO$_2$ detected on the Saturnian
    satellites with the icy Galilean satellites. 
    Dashed lines are shown at 2.6975 and 2.708
    $\mu$m and at 4.25 and 4.27 $\mu$m for reference.
    Ganymede and Callisto have no JWST observations
    at 2.7 $\mu$m with which to compare.}
\end{figure}

These observations have interesting implications for the icy Galilean satellites and the
state of their CO$_2$ as well (Figure 6).
Interpretations for the
CO$_2$ detected on the Galilean satellites 
are sometimes similar to the interpretations
we have made here for the Saturnian satellites,
though in some cases the similarity of the
interpretation is in spite of large spectral
differences.
The CO$_2$ signature on Europa appears intriguingly similar to that on Hyperion and on the trailing
hemisphere of Iapetus. All three bodies exhibit the $\nu_3$ band as a doublet, with wavelengths of approximately
4.25 and 4.27 $\mu$m, and a narrow $\nu_1+\nu_3$ feature near 2.7 $\mu$m. 
On the Saturnian satellites, the differences between Phoebe, Iapetus, and Hyperion
demonstrate that the longer wavelength $\nu_3$ feature is associated with the dark,
plausibly organic
material, while the shorter $\nu_3$ peak and the narrow $\nu_1+\nu_3$ band are associated with the water ice (or at least the icy regions).
On Europa, however, the 2.7 $\mu$m combination band appears more spatially correlated with the longer-wavelength $\nu_3$ peak \citep{villanueva_endogenous_2023}, suggesting that a different interpretation is necessary for these very similar signatures. Indeed, Europa shows no spectral signatures of organic material or evidence of Phoebe-like dark material \citep{trumbo_distribution_2023}.
No satisfactory explanation for the spectral signatures of
CO$_2$ on Europa has yet to be presented, and the resemblance to Iapetus and Hyperion remains a mystery,
though upcoming additional spatial coverage of Europa by JWST may help to resolve some
of these ambiguities.
On Ganymede,
the CO$_2$ at the poles has been interpreted to be trapped in amorphous 
water ice, and the 4.27 $\mu$m band appears similar to that of the inner Saturnian satellites \citep{bockelee-morvan_composition_2024}.
Unfortunately, no shorter wavelength (G235H) JWST NIRSpec data of Ganymede exist, so no comparison to the
2.7 $\mu$m region can be made. 
The equatorial CO$_2$ on Ganymede resembles none of the features on Saturn nor on the other
Galilean satellites and remains a mystery.
On Callisto the $\nu_3$ CO$_2$ band is similar both to that associated with the dark
material on the trailing hemispheres of Dione and Rhea (which has no associated
2.7 $\mu$m absorption) and to that associated with the water ice on the outer
Saturnian satellites (which has a narrow 2.7 $\mu$m absorption). \citet{cartwright_revealing_2024} suggest that, at least on the trailing hemisphere,
the CO$_2$ forms radiolytically in and is trapped by carbonaceous material, though laboratory
studies of the irradiation of organic materials show longer-wavelength $\nu_3$ bands
\citep[i.e.][]{gomis_co2_2005}. The similarity of the $\nu_3$ absorptions on the dark
terrains of Callisto to that of the CO$_2$ associated with the dark contaminants of the inner Saturnian satellites is
intriguing, and, again, a comparison of the 2.7 $\mu$m regions would be instructive, though, again, no 2.7 $\mu$m data for Callisto have yet been
obtained. On the
leading hemisphere of Callisto, the $\nu_3$ band center shifts
in a way consistent with the inclusion
of a CO$_2$ component mixed in water ice, reminiscent of the longer wavelength
$\nu_3$ on the inner satellites of Saturn where a broad 2.71 $\mu$m absorption is also present.

Observations of the Saturnian system, and a comparison across satellites and
across wavelengths, have finally allowed a consistent picture of the sources of
CO$_2$ and the mechanisms of its trapping to begin to be assembled. 
Significant work remains, however, to fully understand CO$_2$ here and elsewhere in
the solar system, including
additional laboratory work exploring trapping materials and mechanisms proposed here, 
a more complete set of observations of the Galilean satellites,
and a more in depth examination of the dark material on the outer satellites 
of the Saturnian system.

\begin{acknowledgments}
We thank two anonymous referees for helping us to strengthen the arguments put in this paper. This work is based on observations made with the NASA/ESA/CSA James Webb Space Telescope. The data were obtained from the Mikulski Archive for Space Telescopes (MAST) at the Space Telescope Science Institute, which is operated by the Association of Universities for Research in Astronomy, Inc., under NASA contract NAS 5-03127 for JWST. These observations are associated with program \#3716. Support for program \#3716 was provided by NASA through a grant from the Space Telescope Science Institute, which is operated by the Association of Universities for Research in Astronomy, Inc., under NASA contract NAS 5-03127. The specific observations analyzed can be accessed via \dataset[ DOI:10.17909/30px-vq56 ]{http:doi.org/10.17909/30px-vq56}
\end{acknowledgments}
\facilities{JWST/NIRSpec.}

%\bibliography{allrefs}
%\bibliographystyle{aasjournal}

\end{document}